\documentclass[letterpaper,11pt]{article}
\usepackage{achemso}
\usepackage{amssymb}
\usepackage{amsmath}
\usepackage{amsfonts}
\usepackage{graphicx}

\newcommand{\wb}{\nu_{\mathrm{b}}}
\newcommand{\wo}{\omega_{\mathrm{o}}}
\newcommand{\w}{\omega}

\newcommand{\nn}{{\vec n}}
\newcommand{\nnp}{{\vec n'}}

\renewcommand{\d}{\mathrm{d}}
\newcommand{\R}{\mathrm{R}}
\newcommand{\middlefig}{.55\textwidth}
\newcommand{\singlefig}{.7\textwidth}
\newcommand{\halffig}{.45\textwidth}
\newcommand{\doublefig}{.9\textwidth}%
\newcommand{\kb}{\mathrm{k}_\mathrm{B}}

\newcommand{\hw}{\hbar\,\omega}
\newcommand{\bhw}{\beta \hbar\,\omega}
\newcommand{\br}{\mathrm{b}} 

\newcommand{\fracc}[2]{\ensuremath \frac{\displaystyle #1}{\displaystyle #2}}
\newcommand{\Ea}{E_\mathrm{a}}

\begin{document}

\title{Discrete breathers for understanding reconstructive mineral
processes at low temperatures}

\author{JFR Archilla\thanks{Corresponding author. Email: archilla@us.es}, J Cuevas\\
Grupo de F\'{\i}sica No Lineal.  Universidad de Sevilla. \\ Departamento de F\'{\i}sica
Aplicada I. ETSI Inform\'{a}tica \\
 Avda. Reina Mercedes, s/n.
41012-Sevilla, Spain
\and MD Alba, M Naranjo and JM Trillo\\
Departamento de Qu\'{\i}mica
Inorg\'anica.  Universidad de Sevilla.\\Instituto de Ciencia de Materiales de Sevilla. \\
Consejo Superior de Investigaciones
Cient\'{i}ficas. \\ P.O. Box 874, 41080-Sevilla (Spain)}

\date{September 20, 2006}
\maketitle

\begin{abstract}
Reconstructive transformations in layered silicates need a high
temperature in order to be observed. However, very recently, some
systems have been found where transformation can be studied at
temperatures 600$^\circ$C below the lowest experimental results previously reported,
including sol-gel methods.
 We explore the
possible relation with the existence of intrinsic localized modes,
known as discrete breathers. We construct a model for nonlinear vibrations within
the cation layer, obtain their parameters and calculate them numerically, obtaining
their energies. Their statistics shows that although there are far less  breathers
than phonons, there are much more above the activation energy, being therefore a good
candidate to explain the reconstructive transformations at low temperature.
\\ \mbox{}
\end{abstract}


\section{Introduction}
During the last decade, some of the
present authors have achieved the synthesis of crystalline
high-temperature polymorphs of rare earth (RE) disilicates
(RE$_2$Si$_2$O$_7$) at non-expected temperatures, significantly
lower than those previously reported, through a reconstructive
process (LTRT) from clay minerals as the silicon
source~\cite{JPC94,JPC96,AmMin01a,Becerro03}.

This finding is of general importance in the development of
advanced structural ceramics~\cite{HYISNM02}
 or the storage of
radioactive wastes~\cite{AmMin01a} and should allow completion of the
available Si~O$_2$-RE$_2$O$_3$ phase diagram~\cite{F73}.
  Although no precise
explanation has been found up to now, some of the present authors
had previously suggested a {\em chimie douce} mechanism based on
the diffusion of RE ions into the interlayer space of the
expandable clay minerals~\cite{AmMin01a,Becerro03}.

MacKay and Aubry~\cite{MA94} have suggested that a possible effect
of the existence of localized nonlinear vibrations, named discrete
breathers (DBs) could be an apparent violation of Arrhenius' law,
i.e., the phenomenon of chemical reactions taking place at much
lower temperatures than expected. Although this hypothesis is
adventurous, it is worth exploring. Moreover, experimental evidence
of DBs has already been found in several systems such as
antiferromagnets~\cite{SES99}, waveguide arrays~\cite{FCSEC03},
molecular crystals~\cite{Swa} or Josephson-junctions~\cite{TMO00}.
  Moving DBs have been proposed as an explanation
 of dark tracks in muscovite~\cite{MER98}
 and there existence in muscovite has just been proven through a sputtering
  experiment~\cite{RE06}.

With this aim, we have made calculations and shown
 that the contribution of DBs can provide an
interpretation for LTRT in clay minerals. And in order to give
 experimental support to the hypothesis of DBs, it will be
shown that the LTRT phenomenon is not exclusive of expandable clay
minerals, as expected by the previously suggested "chimie douce"
mechanism~\cite{AmMin01a,Becerro03}, but also extensible to
non-expandable layered silicates, such as mica muscovite. The
layout of this work consists of: Section~2: Some structural
considerations on the reconstructive nature of transformation of
layered silicates into disilicates; Section~3: A report on a new
experiment on LTRT performed by the authors on mica muscovite;
Section~4: Argumentation about the difficulty to explain by the
conventional chemical kinetics model the latter experiment; Section~5:
Description of an alternative model based on DBs with numerical
calculations; Section~6: After a summary of breather
statistics theory, the description of our numerical simulations
and the consequences on the reaction rate, we
conclude with the possibility of explaining, for the
first time, the LTRT phenomenon in the synthesis of
high-temperature polymorphs of silicates by the contribution of
DBs. The article in itself is ended with a summary. Two
appendices give some detail on phonon and breather statistics
respectively.

\section{From layered silicates to disilicate crystal
structures}
 The synthesis of disilicates from layered silicates,
expandable as clay minerals or non-expandable as mica, actually
means a reconstructive transformation as shown below.

Layered silicates are made up from two basic building blocks: a
sheet of edge-sharing [SiO$_4$] units, the tetrahedral sheet, and
another one of edge-sharing [MO$_6$], the octahedral sheet. There
are three main groups of layered silicate minerals, according to
the combinations of tetrahedral and octahedral sheets: 1:1, 2:1
and 2:1:1. In 2:1, one octahedral sheet is sandwiched between the
apices of two tetrahedral ones. In these, also called T-O-T
silicates, layers are either held together by weak van der Waals
forces if they are neutral, or may have cations between them for
charge balance if substitutions in either tetrahedral or
octahedral sheet result in a net layer charge. The 2:1 layered
silicates are classified as trioctahedral or dioctahedral, after
the full occupation of the octahedral sheet by Mg(II) or two
thirds by Al(III). Talc (trioctahedral) and pyrophillite
(dioctahedral) are minerals with non-charged layers. In the case
of low charge, it results that the clay minerals have the
capacity to expand by taking up H$_2$O molecules in the
interlayer space. For high charge, there is mica: phlogopite
(trioctahedral) and muscovite (dioctahedral).

\begin{figure}[t] \begin{center}
    \includegraphics[width=\halffig]{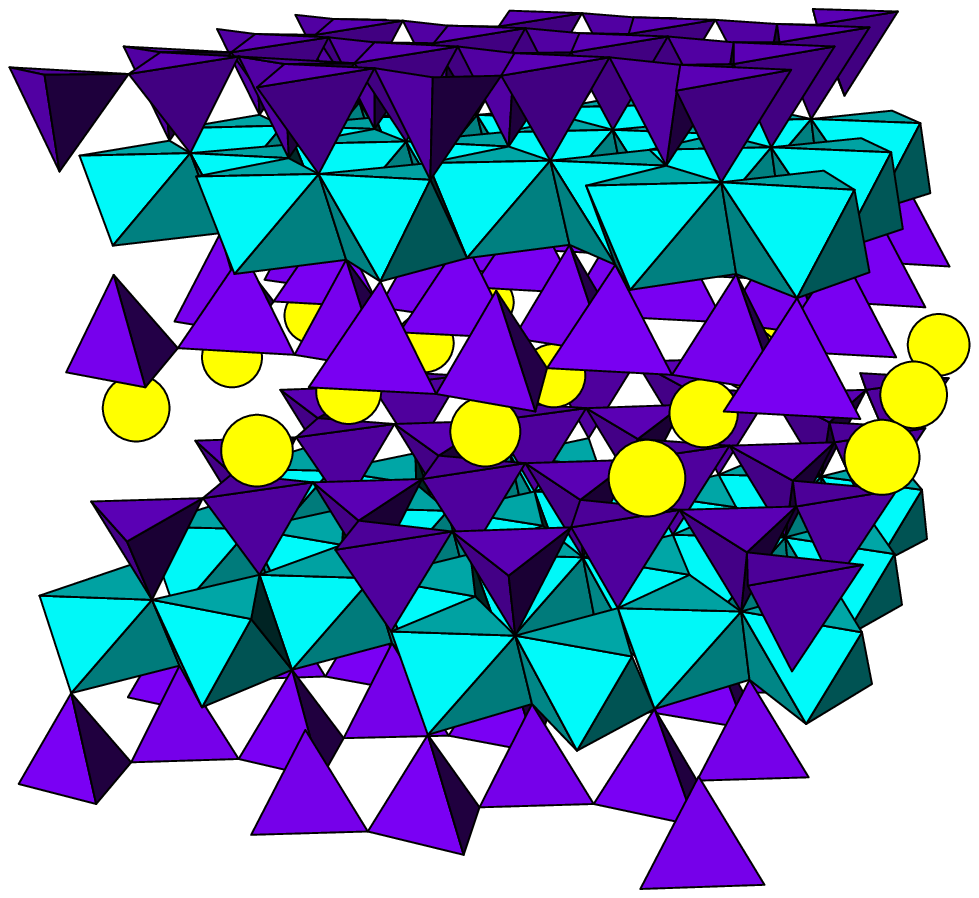}\quad
\includegraphics[width=\halffig]{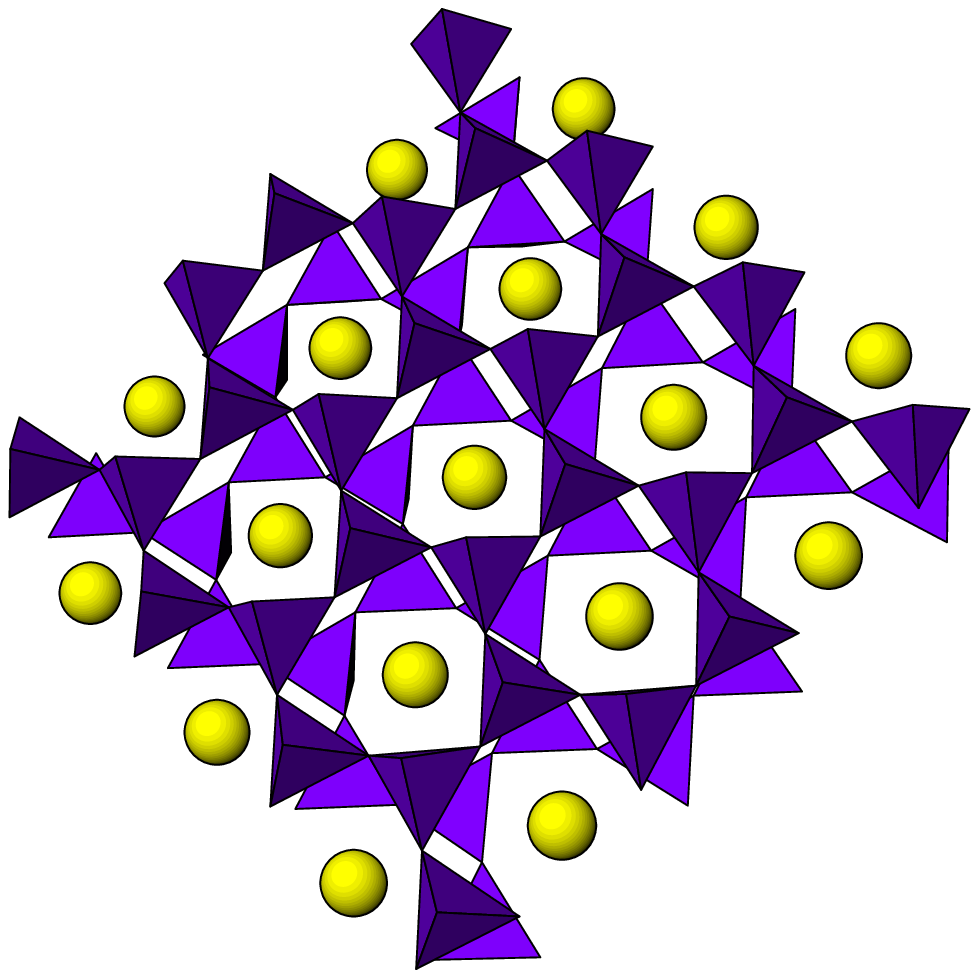} \caption{Crystal
unit cell of muscovite ICSD 34406. The circles represent the
potassium ions forming the interlayer  sheet.
($\textrm{a}=5.19$
\AA; $\textrm{b}=9.02$ \AA; $\textrm{c}=20.0$ \AA;
$\beta$=95.7$^\circ$) } \label{fig:unitcell} \end{center}
\end{figure}

 Muscovite is a mica which layer charge
comes from the isomorphic substitution of silicon by aluminium in
the tetrahedral sheet~\cite{PFAPP03}.
 The potassium located in the
interlayer space, for charge balance, cannot be hydrated; thus,
muscovite does not expand. Its structure is depicted at the left of
Fig.~\ref{fig:unitcell}. At its right, the interlayer space of
muscovite is illustrated and it can be observed that both surfaces
of the upper and lower tetrahedral sheet are formed by the basal
oxygen atoms from the [SiO$_4$] tetrahedra, which form a rough
hexagonal honeycomb structure. The interlayer balancing sheet is
therefore sandwiched in between, potassium occupying the dimples
left at the centre of each pair of hexagonal cells. In real
crystals, although the on-site potential created by the silicate
layer tends to preserve the symmetry, there are always distortions
like tetrahedral rotation (the actual situation in muscovite is
shown in this figure). Each potassium ion is surrounded by six other
potassium ions in the interlayer sheet with a ditrigonal symmetry.
The distance between the potassium and the basal oxygens layers is
1.45 \AA.

\begin{figure}[t] \begin{center}
    \includegraphics[width=\middlefig]{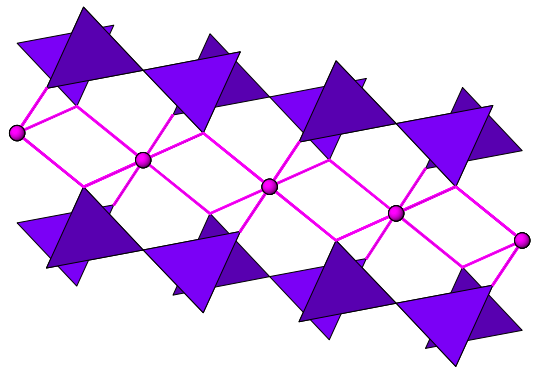}
\caption{Structure of lutetium disilicate. The circles represent
the lutetium ions.}
\label{fig:disili} \end{center} \end{figure}

 Essentially, in both clay minerals and mica [SiO$_4$]
tetrahedra are linked into infinite two-dimensional networks by
sharing three oxygens. However, disilicates, or pyrosilicates, are
the simplest of the condensed forms, where only two tetrahedra
share one edge and constitutes the anion Si$_2$O$_7$$^{6-}$
 (Fig~\ref{fig:disili}).
 The transformation of any layered-silicate into disilicate
involves the rupture of two silicon-oxygen bonds by each
[SiO$_4$], whatever the reaction mechanism might be, the
transformation being reconstructive.

It is well-known that natural pyrosilicates show a wide range of
Si-O-Si angle, from 131$^\circ$ to 180$^\circ$~\cite{Corton}, and
that the activation energy is reduced if the surface and strain
energy terms are diminished by good lattice matching across the
interface between the new and parent phase. However, this is not
the present case. It rather seems that the disruption of the
tetrahedral sheet could be the consequence of localized nonlinear
vibration modes as commented in the introduction and explained in
the following sections.

\section{Experimental RE-disilicates synthesis}

The method used by us to synthesize RE-disilicates consists of a
hydrothermal reconstructive process at low temperatures in an
isolated reaction vessel constructed in our laboratory. A layered
silicate and an aqueous solution are the silicon and RE(III)
sources, respectively. Up to now, a set of expandable clay minerals
had been studied, rendering conclusions on the relationship between
mineralogical compositions and reactivity~\cite{AmMin01a}. The reaction
temperatures were always below the critical one of water; thus both
vapour and liquid phases coexist throughout the whole reaction.
Reconstructive structural changes occurring in the layered silicate
are always analyzed studying the long-range order by X-ray powder
diffraction (XRD), the chemical environment of the main constituent
elements of the lattice by magic-angle spinning nuclear magnetic
resonance spectroscopy (MAS-NMR) and the microstructural and
microchemical composition by electron microscopy (SEM) and by
energy-dispersive X-ray (EDX).

As it has been already mentioned, it has been suggested~\cite{MA94}
that DBs might bring about an apparent violation of Arrhenius law,
leading to chemical reactions being observable at much lower
temperatures than expected~\cite{AmMin01a}. It has also been suggested
that DBs in the interlayer potassium sheet may be responsible for
the dark lines observed in crystals of mica
muscovite~\cite{SRV93,MER98} and the existence of moving breathers in mica has
been recently proven through a sputtering experiment~\cite{RE06}.
 It implies a LTRT process occurring in a
nonexpendable layered silicate, which contradicts the mechanism
published by some of the present authors to explain the synthesis of
RE-disilicates as associated to the capacity for
expanding~\cite{AmMin01a}. In order to confirm experimentally the
hypothesis that the LTRT phenomenon is not an exclusive feature of clay
minerals, but can also be attributed to non-expandable layered silicates, we
have performed the hydrothermal synthesis of Lu-disilicate from
muscovite, under the same experimental conditions as those used for clay
minerals.

\begin{figure}[ht] \begin{center}
    \includegraphics[clip,width=\singlefig]{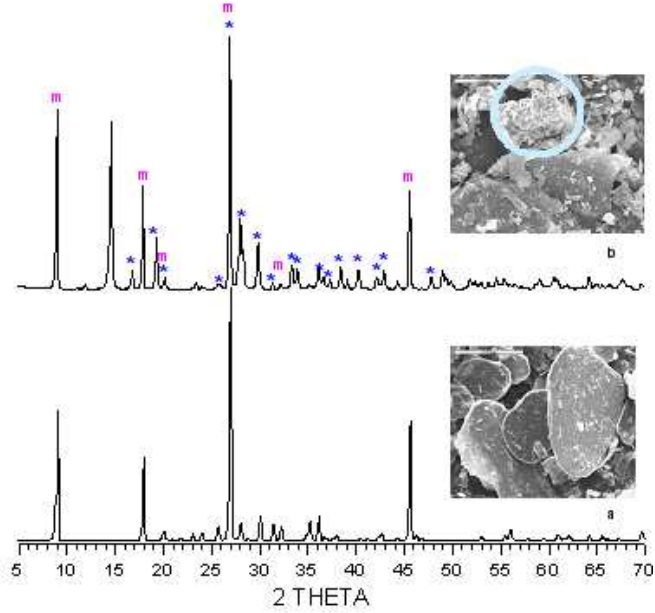}
\caption{XRD pattern and SEM micrography of untreated (a) and
hydrothermally treated (b) muscovite. m=muscovite,
*=Lu$_2$Si$_2$O$_7$. The composition of the rough particle inside
the circle is compatible with Lu$_2$Si$_2$O$_7$}
\label{fig:XRD}
\end{center} \end{figure}

When muscovite is hydrothermally treated in stainless steel reactors
at 300 $^\circ$C for 72 hours with lutetium nitrate 0.05 M solution,
it gives rise to Lu$_2$Si$_2$O$_7$. Figure~\ref{fig:XRD}a  shows the
XRD diagram of the untreated muscovite. It reveals numerous hkl
basal reflections compatible with the 2M$_1$ polytype and a perfect
ordering of the layers. After the hydrothermal treatment, the XRD
pattern shown in figure~\ref{fig:XRD}b exhibits a number of specific
reflections which are
consistent with the development of a new crystalline phase
Lu$_2$Si$_2$O$_7$ (JCPDS file number 76-1871). The SEM photograph of
the mica shows big flakes whereas the sample submitted to
hydrothermal treatment, in addition  reveals irregular, rough
particles, which correspond to Lu$_2$Si$_2$O$_7$.
The $^{29}$Si MAS NMR spectra of both expandable and non--expandable
layered silicates show an evolution from Q$^3$ silicon environment
to Q$^1$ environment~\cite{albachain05}.
This result supports
that the LTRT phenomenon is common to expandable and non-expandable
layered silicates, both containing an interlayer sheet of cations to
balance the layer charge.

\begin{figure}[ht] \begin{center}
    \includegraphics[clip,width=\singlefig]{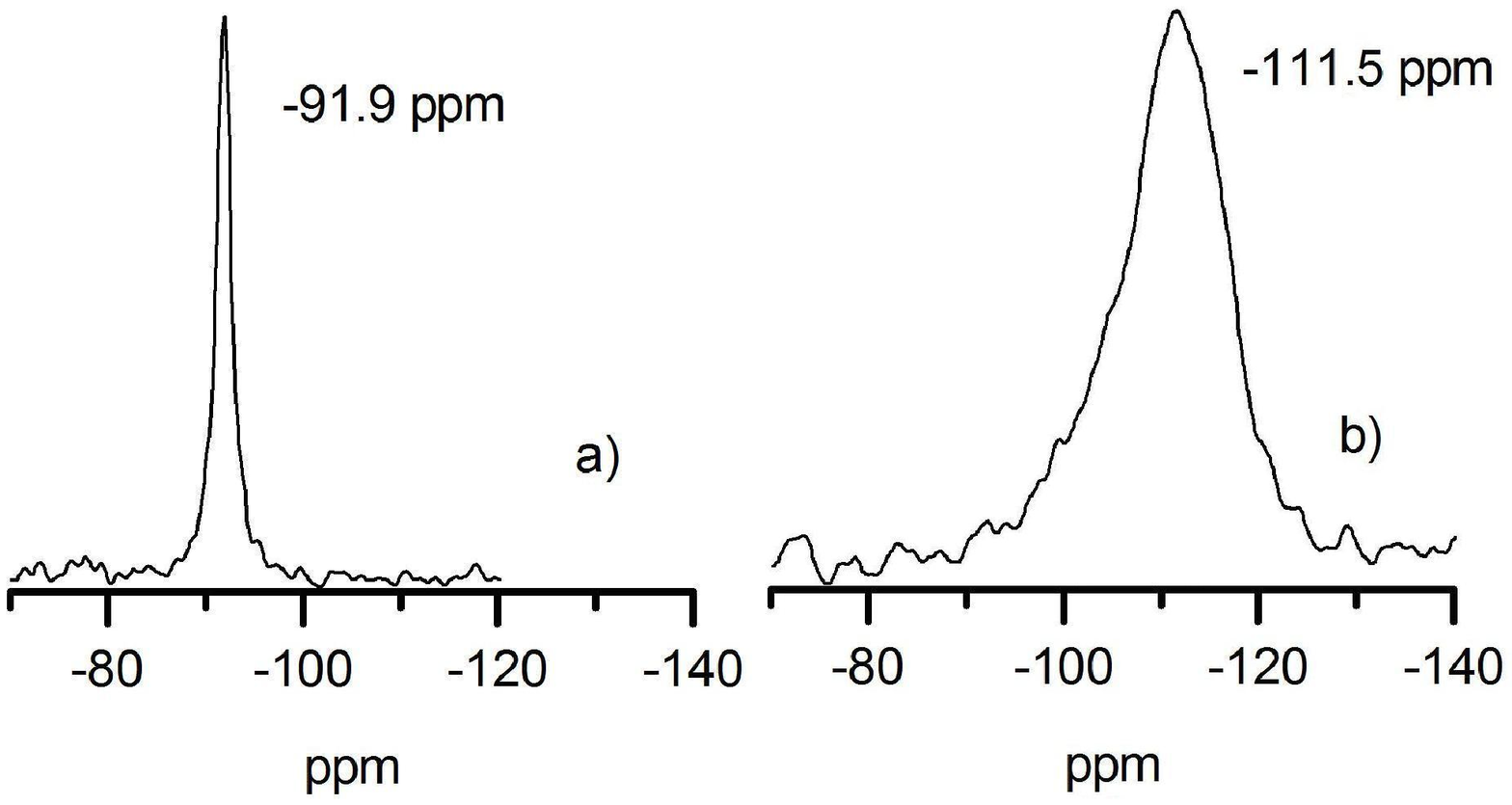}
\caption{$^{29}$Si MAS NMR spectra: a) Lu$_2$Si$_2$O$_7$ b)
SiO$_2$ treated with 50 ml of Lu(NO$_3$)$_3$ at 300~$^\circ$C}
\label{fig:silice}
\end{center} \end{figure}

Note that although the hydrothermal treatment of layered
silicates, muscovite and others, is, effectively, at the origin
of the development of a new crystalline phase, as a necessary
experimental condition, it is not sufficient. In the case of a
silicon source different from a layered silicate, which has been
used by the authors for the first time, the process does not
occur. An example is illustrated in the Fig.~\ref{fig:silice}.
This figure shows the $^{29}$Si MAS NMR of SiO$_2$ submitted to a
hydrothermal treatment at 300~$^\circ$C (b) and the typical
spectrum of the lutetium disilicate (a). The signal of the
SiO$_2$ after hydrothermal treatment remains in the chemical
shift range typical of Q$^4$ environment (typical of SiO$_2$),
which is clearly different for Q$^1$ environment (typical for the
new phase Lu$_2$Si$_2$O$_7$). Thus, it demonstrates that under
hydrothermal condition, SiO$_2$ does not transform in a new phase.

 \section{The conventional chemical kinetics approach}
Transformation processes of minerals in which there is a major
reorganization with bonds and, even, change in the chemical
composition are classified as {\em reconstructive}. These obey
mechanisms which involve very high activation energies when the
rupture of strong bonds are involved. Synthesis of RE-disilicates
from layered silicates requires the rupture  of silicon-oxygen
bonds, which are considerably stronger than the bonds between any other
element and oxygen. Silicate minerals make up the vast majority
of rocks and their reconstructive transformation processes show
activation energies as high as 200 kJ/mol or even higher~\cite{Putnis}.
Therefore, these transformations can be observed in silicate-based
minerals at temperatures higher than approximately 1000~$^\circ$C and
are apparently impossible at lower ones.

 It is well known that the over-all effect on temperature on the reaction rate
 constant $k$ is expressed by Arrhenius law:
\begin{equation} k = A\, \exp(-\Ea/RT)
\end{equation} where $A$ and $\Ea$ are the frequency factor
and the activation energy, respectively. According to it, and
without any further insight into the mechanism, the rate of a
reaction is about $10^9$ times faster at 1000~$^\circ$C than it is
at 300~$^\circ$C for an activation energy of 200 kJ/mol.

It can be shown that the second parameter in the Arrhenius law, namely, the
frequency factor $A$, does not support LTRT synthesis of
RE-disilicates as well.
Usually, reaction rates are described within the frame
of the quasi--equilibrium activated state model~\cite{Moore62}.
According to the latter,  the necessary
condition for a transformation to take place at a measurable rate
is that a sufficient number of atoms have enough energy to achieve the
transition state. This energy is supplied by thermal fluctuations.
The rate of reaction is then simply the number of activated complexes
passing per second over the potential barrier.

Applying the conventional transition state (activated complex)
theory~\cite{Eyring,Back,Pelzer,Evans,Eyring2,Wynne} to our case, the simplest
formulation of the mechanism can be cast in the following form:
 \begin{eqnarray}
 \textrm{S (layered silicate) +RE(III)(aq)} &\leftrightarrow &
 \textrm{[S-RE(III)]$^*$} \nonumber\\
\textrm{[S-RE(III)]$^*$}&\rightarrow & \textrm{RE$_2$Si$_2$0$_7$}
\end{eqnarray}
 The rate constant $k$ for the reaction can be
derived by assuming that the transition state (or activated
complex) is in equilibrium with the reactants. If C$^*$
represents the concentration of the transition state then the
equilibrium constant is:
 \begin{equation}
 K^*=\frac{\textrm{C}^*}{\textrm{[RE(III)]}}
 \end{equation}
The rate constant $k$ and the equilibrium constant $K^*$ are
related by the Eyring equation:
 \begin{equation}
k=\nu\, K^*\,,
 \end{equation}
 with $\nu=\mathrm{k}_\mathrm{B}\,T/\mathrm{h}$, h being the Planck constant.

To be precise, the rhs of the above expression should be
multiplied by a factor $\eta$, the transmission coefficient,
which is the probability that the complex will dissociate into
products instead of back into reactants. For most reactions $\eta$
is between 0.5 and 1.0~\cite{Moore62}. Through a thermodynamic
formulation of $K^*$, it results:
 \begin{eqnarray}
k &=& \nu \,\eta \exp (-\Delta G^{0*}/\R T) = \nonumber \\
 & &\nu \,\eta  \exp (\Delta S^{0*}/\R) \exp (-\Delta H^{0*}/\R T) =
 A \exp (-\Ea/R T )\,,
  \end{eqnarray}
 where the superindex $^0$ stands for normal conditions.
  In liquid and solid
systems, the $p\Delta V^{0 *}$ term is negligible and $\Delta H^0
= \Delta E^{0*} = \Ea$.

At 300$^\circ$C we have calculated the factor $A$ by substitution
of all the parameters in the above expression and it takes the
usual value $10^{13}-10^{14}$ s$^{-1}$ for a first order
reaction. It does not explain the observation of LTRT synthesis
of RE-disilicates as previously concluded from $\Ea$.

It is well known that, in parallel to the reorganization of the clay,
there must be  nucleation of RE$_2$Si$_2$O$_7$ crystals and that the
activation energy is reduced if the surface and strain energy
terms are diminished by good lattice matching across the
interface between the new and parent phase~\cite{Becerro03}.
However, this is not the present case. It rather seems that the
disruption of the tetrahedral sheet is the consequence of
localized nonlinear vibration modes, as suggested in
Ref.~\citenum{MA94}. If the vibration modes were delocalized, the
relationship shown in Ref.~\citenum{Putnis} between bond angle, Si-O
distance and free energy would be incompatible with an
appreciable parent structure-directing character.

Ytrium disilicate (Y$_2$Si$_2$O$_7$) has four polymorphs, namely $y$,
$\beta$,
$\gamma$ and $\delta$. Our LTRT synthesis from the two layered silicates
saponite and laponite have shown a structure-directing character of
the parent clay by only giving $y$-Y$_2$Si$_2$O$_7$ and
$\delta$-Y$_2$Si$_2$O$_7$, the lower and higher temperature
polymorphs~\cite{Becerro03}. The relative position of the two
tetrahedra in the disilicate unit of the $y$- and $\gamma$- polymorphs
are similar to their position in the tetrahedral sheet of saponite and laponite,
which is $\sim 141^{\circ} $. The Si-O-Si bond angles in Y$_2$Si$_2$O$_7$
polymorphs are 134$^{\circ}$ ($y$), 180$^{\circ}$ ($\beta$),
170$^{\circ}$ ($\gamma$) and  158$^{\circ}$ ($\delta$). The percentages
of $y$- and $\delta$- phases in the cases of saponite and laponite
are explained in Ref.~\citenum{Becerro03} as related with the presence
of Al(III) in the precursor framework. Moreover, the importance of
maintaining the local Si-O-Si bond angle of
 precursor structure is shown by the fact that the
$\delta$--polymorph has been synthesized at more than 365$^\circ$C below
the stability range shown in the phase diagram:
 \begin{eqnarray}
 \textrm{y-Y$_2$Si$_2$O$_7$} \rightarrow
 \beta\textrm{-Y$_2$Si$_2$O$_7$}&\mbox{}& {1050(50)~^\circ\textrm{C}}\nonumber \\
\beta\textrm{-Y$_2$Si$_2$O$_7$} \rightarrow
 \gamma\textrm{-Y$_2$Si$_2$O$_7$}&\mbox{}&{1350(50)~^\circ\textrm{C}}\nonumber \\
 \beta\textrm{-Y$_2$Si$_2$O$_7$} \rightarrow
 \delta\textrm{-Y$_2$Si$_2$O$_7$}&\mbox{}&{1500(50)~^\circ\textrm{C}}\nonumber
\end{eqnarray}

At the present stage of DBs application to understanding the LTRT
phenomenon, other experimental results already published by
our group, such as the influence of isomorphic substitution in clay
minerals on the reactivity of the latter, are still awaiting their
explanation.
 We
hope that the further work on DB in solid state physics will permit to
interpret all those experimental results in the near future.

 \section{Discrete breathers model for potassium vibrations}
\label{sec:model}
  As commented above, it has been
predicted that discrete breathers could bring about an increase of
the reaction rate~\cite{MA94}. They are localized vibrational
modes in networks of nonlinear oscillators, which have been
thoroughly studied in the last years~\cite{FW98,PHYSD99,CHAOS03}.
They have been observed in
experiments~\cite{SES99,FCSEC03,Swa,TMO00} and are thought to play
an important role in DNA denaturation~\cite{P04}. They were  also
suggested to be responsible for dark lines in crystals of
muscovite~\cite{SRV93,MER98} and their existence demonstrated in muscovite
through a sputtering experiment~\cite{RE06}.

\begin{figure}[ht] \begin{center}
    \includegraphics[clip,width=\middlefig]{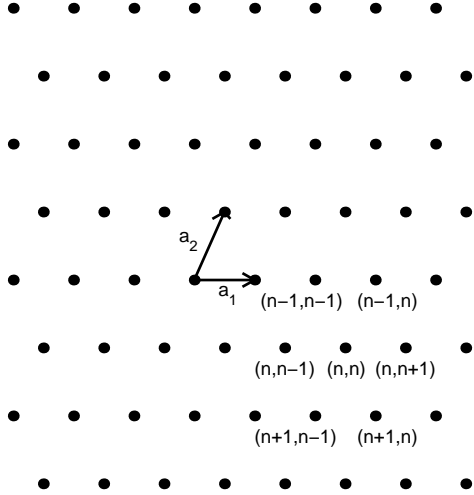}
\caption{Scheme of the 2D lattice used for discrete breather
calculation. a$_1$ and a$_2$ are the base vectors.}
\label{fig:hex} \end{center} \end{figure}

In this paper we have considered breathers for the out-of-plane
movements of the potassium cations, i.e. the movements transversal
to the cation layer. This layer is considered to be a 2D
triangular lattice, as Fig.~\ref{fig:hex} shows. The Hamiltonian
is given by:
\begin{equation}\label{eq:Ham}
    H=\sum_\nn\left(\frac{1}{2}\,m\dot u_\nn^2 + V(u_\nn) + \frac{1}{2}\,
    \kappa \sum_{\nnp\in NN}(u_\nn-u_\nnp)^2\right),
\end{equation}
 where $m=39.1\,\mathrm{amu}$ is the mass of a
potassium cation, $\kappa$ is the elastic constant of the
cation-cation bond, $V(u_\nn)$ is an on--site potential, and the
second sum is extended to the nearest-neighbours, as indicated in
Fig.~\ref{fig:hex}. The value of $\kappa$, i.e. the elastic
constant for potassium-potassium bond stretching is $10\pm1$ N/m,
after Ref.~\citenum{HCP}.

\begin{figure}[ht]
\begin{center}
    \includegraphics[clip,width=\singlefig]{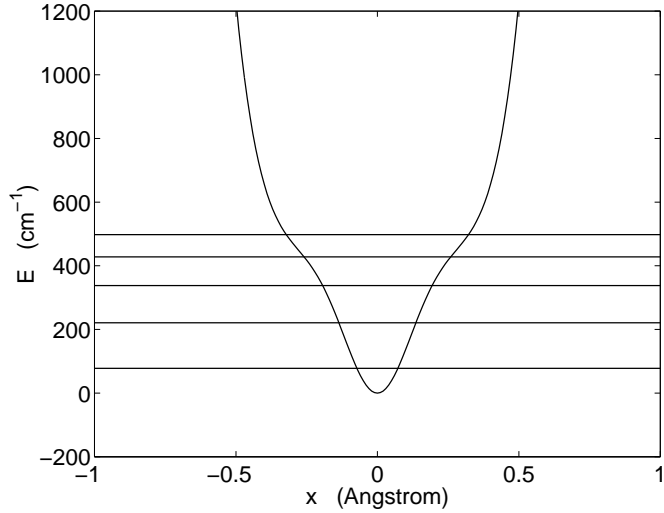}
\caption{ Numerically calculated on-site potential for the
potassium vibration based in the observed bands (horizontal lines)
} \label{fig:potmica} \end{center}
 \end{figure}

A band at 143~cm$^{-1}$ is identified in Ref.~\citenum{Farmer} with
the K$^+$ vibration perpendicular to the K$^+$-plane in spectra
from 30 to 230~cm$^{-1}$. We have performed infrared spectra in
CNRS-LADIR~\footnote{Laboratoire de Dynamique, Interactions et
R\'eactivit\'e at CNRS-Thiais, Paris } above 200~cm$^{-1}$. There appear
bands at 260, 350 and 420~cm$^{-1}$ which we assign tentatively
to higher order transitions of the same vibration.
 Using standard numerical methods to solve the Schr\"odinger equation for
 the K$^+$ vibrations with a
 potential composed of the linear combination with three gaussians and a polynomial of
 degree six, with adjustable parameters, we have been able to find a suitable potential
 that fits these bands and their intensities. It is given by:
 \begin{equation}
    V(u)=D(1-\exp(-b^2u^2))+\gamma u^6,
 \end{equation}
 with $D=453.11$ cm$^{-1}$, $b^2=36.0023$
\AA$^{-2}$ and $\gamma=49884$ cm$^{-1}$\AA$^{-6}$.
To completely define the potential, the limitation to
 the  K$^+$ displacement due to the muscovite structure has also
 been taken into account.
 Fig.~\ref{fig:potmica} shows the adjusted potential together
with the observed bands.

The nonlinearity of the on-site potential allows the existence of
intrinsic localized modes or \emph{discrete breathers} apart from
low-amplitude linear modes~\cite{MA94}. These localized solutions
exist as long as no integer multiple of their frequency resonates
with the frequency of a linear mode and
can be obtained numerically with machine precision. They are calculated
using procedures based on the anti-continuous  limit ($\kappa=0$)
\cite{MA96}, which consists of calculating an orbit of an
isolated oscillator submitted to the potential $V(u)$ with a
fixed frequency $\wb$, and using this solution as a seed for
calculating the solution of the full dynamical equations
($\kappa\neq0$) through a continuation method.

In principle, we have considered
excitations in the soft part of the on-site potential (the energy
of an isolated oscillator decreases with the frequency) because this is
the region that fits the experimental data. In this case, the
frequency of a breather $\wb$ is below the linear modes band and
none of its harmonics coincide with the frequency of a linear
mode for a given value of $\kappa$.

The linear modes are solutions of the equations:
\begin{eqnarray}
 m\ddot  u_{n,n'}+m\,\wo^2 u_{n,n'}
    -\kappa \,( u_{n,n'-1}+
    u_{n,n'+1}+ u_{n-1,n'-1}+
    u_{n-1,n'}+\nonumber \\u_{n+1,n'-1}+ u_{n+1,n'}-6u_{n,n'})=0\,,
\end{eqnarray}

with $\wo^2=V''(0)/m$. Then, the linear modes spectrum is the
following:

\begin{equation}
\w^2=\wo^2[1+4\,C\,(\sin^2(\theta_1/2)
+\sin^2(\theta_2/2)+\sin^2(\theta_2/2-\theta_1/2))],
\end{equation}

with $C=\kappa/(m\,\wo^2)\approx0.15$ and $\theta_i\in[-\pi,\pi]$.
Thus, the frequency of the linear modes lies in the interval
$\nu\in\nu_0(1,\sqrt{1+9\,C})$, with $\nu_0=\wo/(2\pi)=167.50$
cm$^{-1}$. In consequence, no multiple of $\wb$ must lie in this
band.

\begin{figure}[ht]
\begin{center}
    \includegraphics[clip,width=\halffig]{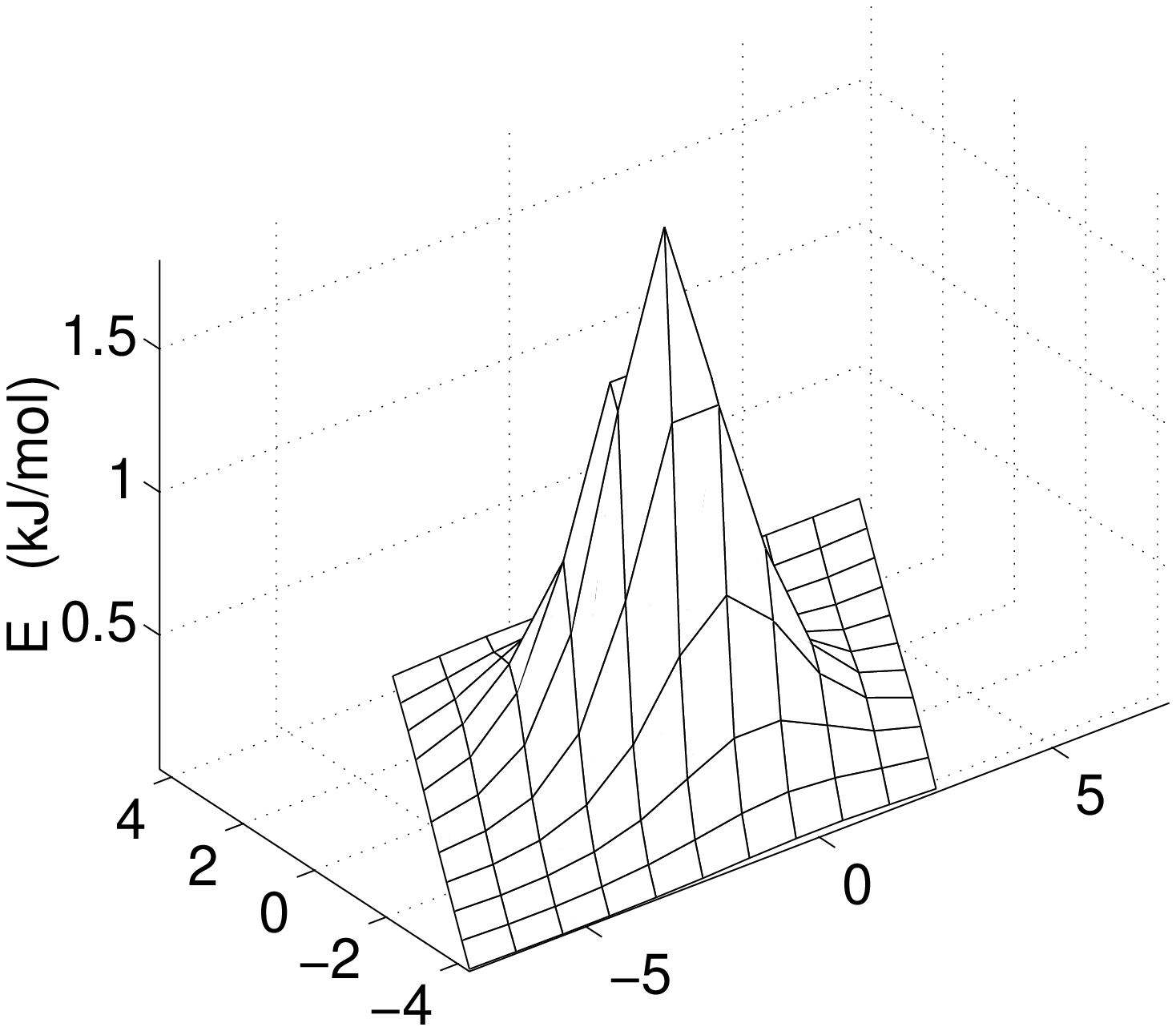}
    \includegraphics[clip,width=\halffig]{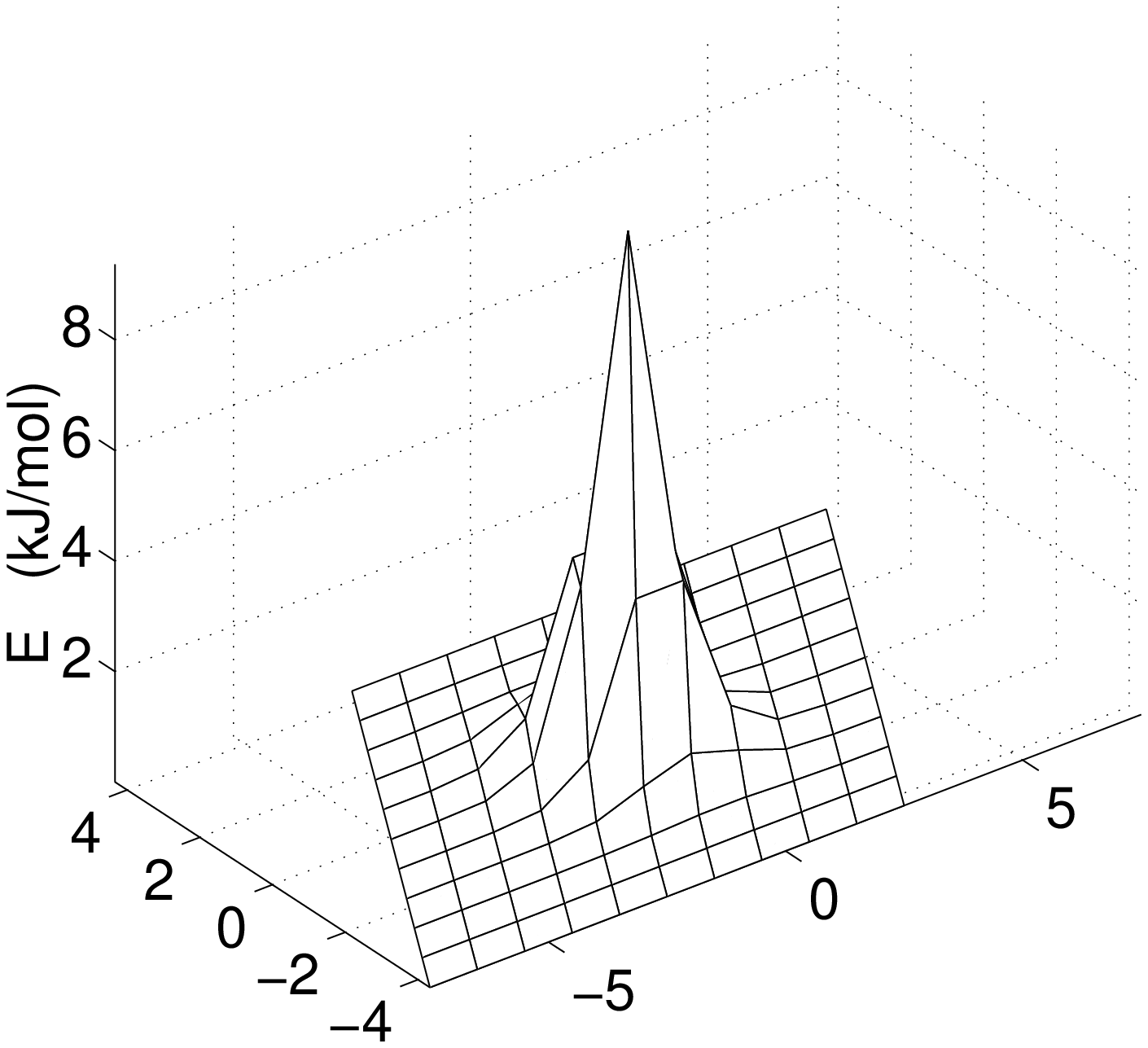}
\caption{(Left)
Energy density profile for a breather with
frequency $\nu_\br=0.97\,\nu_0$ and energy $25.6$~kJ/mol.
(Right) The same but for a breather with frequency
$\nu_\br =0.85\,\nu_0$ and energy $36.3$~kJ/mol. Notice
that the localization is higher for the lower frequency.
The $x$ and $y$ coordinates are in lattice units of $5.2$~\AA,
$\nu_0=167.5$~cm$^{-1}$.}
\label{fig:breather} \end{center}
 \end{figure}

\begin{figure}[ht] \begin{center}
    \includegraphics[clip,width=\singlefig]{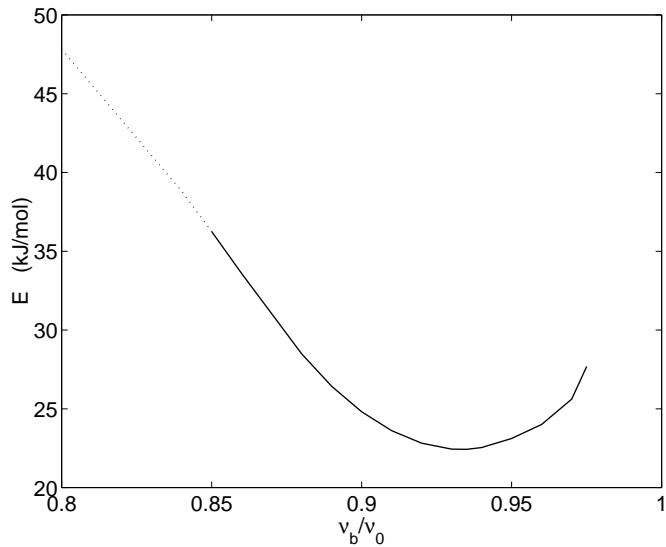}
\caption{Dependence of the breather energies
with the relative frequency for
symmetric single breathers
with frequencies below the phonon band. The dashed line indicates that breathers are unstable.
$\nu_0=167.5$~cm$^{-1}$.}
 \label{fig:energy} \end{center} \end{figure}

Fig.~\ref{fig:breather} shows the profiles of two breathers with
different frequencies and Fig.~\ref{fig:energy} shows the
dependence of the breather energy with respect to the frequency.
We can observe that there is a minimum energy $\Delta\sim
23\,\mathrm{kJ/mol}$ for breather creation as it is known to
happen for 2-D and 3-D breathers~\cite{FKM97}.

These breathers are single symmetric breathers, but there are many
other different types, as breathers with different symmetries, multibreathers,
breathers above the phonon band, which
will have different energies, range of existence and probability
of excitation. For example, symmetric single breathers with frequencies
above the phonon band have energies between 240 and 500 kJ/mol.
The obtention of a complete description for all the
different breather types and their
ranges of existence and stability is a long task that we do not pursue here.

\section{Breather statistics and its effect on the reaction rate}
\label{sec:stat}
\subsection{Breather statistics theory}
\label{subsec:brstat}

 The effect of the temperature on the reaction rate constant $k$
is given by the Arrhenius law $k=A\,\exp(-\beta\,\Ea)$,
with $\beta=1/\R T$.  The Boltzmann factor
$\exp(-\beta\,\Ea)$ represents the fraction of
vibrational modes that are able to deliver the activation energy
$\Ea$ for the reaction to proceed. The frequency
factor $A$ depends on the reaction kinetics and although it is
difficult to calculate the estimation given above is
$10^{13}-10^{14}$~s$^{-1}$, or the order of magnitude of most
chemical reactions.

The statistics of breathers is not at all completed and a full
statistical analysis of muscovite and the reaction kinetics is a
formidable task.  Here we use the theory developed
for 2-D breathers by Piazza {\em et al}~\cite{Piazza03} which is described
in the first subsection of Appendix~\ref{ap:breather}, with
some modifications in subsection~\ref{subsec:brsoft}.
 Note that this theory is not deduced
from first principles but it is based in heuristic
reasoning. It has to be considered a reasonable approximation justified
by its good accuracy with numerical simulations. Here we summarize the
main results that are useful to us:

a)~The probability density $P_\br (E)$, defined so as $P_\br(E)\,\d E$
is the probability of existence or the mean fraction
of breathers with energies between $E$ and $E +\d E$,  is given by
\begin{equation}
P_\br (E)\,\d E=\frac{\beta^{z+1}}{\Gamma(z+1)}(E-\Delta)^z\exp[-\beta (E-\Delta) ]\, ,
\label{eq:pede}
\end{equation}
with $\Gamma(z+1)$, the gamma function, and $z$, a parameter which does not
depend on the energy. The cumulative probability,
$C_\br (E)$ or mean fraction of breathers with energy above
or equal to
$E$ is given by:
\begin{equation}
C_\br(E)=\Gamma(z+1)^{-1}\,\gamma(z+1,\beta\,(E-\Delta))\,,
\label{eq:ce}
\end{equation}
where $\gamma(z,\epsilon)$ is the incomplete gamma function,
defined as $\int_\epsilon^\infty y^{z-1}\exp(-y)\,\d y $~\cite{Abram}.

b)~A population of large breathers tends to develop with mean
energy $\langle E\rangle =\Delta+(z+1)\,\kb T$ and maximum probability for
$E(P_{\rm{max}})=\Delta+ z\,\kb T$.

Fig.~\ref{fig:pe_brph} shows the probabilities and cumulative probabilities of
phonons and breathers, the latter for $\Delta=20$~kJ/mol and $z=2$. Note that
for large energies of the order of the activation energies, $C_\br(E)$, although
small, is
several orders of magnitude larger than $C_{\rm ph}(E)$. The drawback is
that the number of breathers per site is much smaller than the number of phonons, with
typical values of $~10^{-3}$ obtained in numerical simulations.

This theory can be modified to take into account that some breather types
have an upper limit for their energies, due to bifurcations or instabilities as
explained in \ref{subsec:brhard}.
The modified probability densities and cumulative probabilities are
given in Eqs.~(\ref{eq:P_bEm},\ref{eq:CbEm}).

\subsection{Numerical simulations}
\label{subsec:numerical}
We have performed a number of numerical simulations with the following procedure:
a)~A network of $50\times 50$ oscillators
is given a number of times (about 500) the same energy
with different random distributions of velocities; b)~They are left
evolve a sufficiently long time so as they attain states of thermal equilibrium, for which
the mean temperature for all the simulations is calculated; c)~The networks are
cooled by adding a dissipative
term to the dynamical equations for the oscillators at its borders until
the energy stops diminishing;
d)~The number of breathers is counted and their energies are calculated.
The number of breathers divided by the number of sites and the number of simulations
gives the  mean number of breathers $\langle n_\br \rangle$.

\begin{figure}[ht] \begin{center}
    \includegraphics[clip,width=\singlefig]{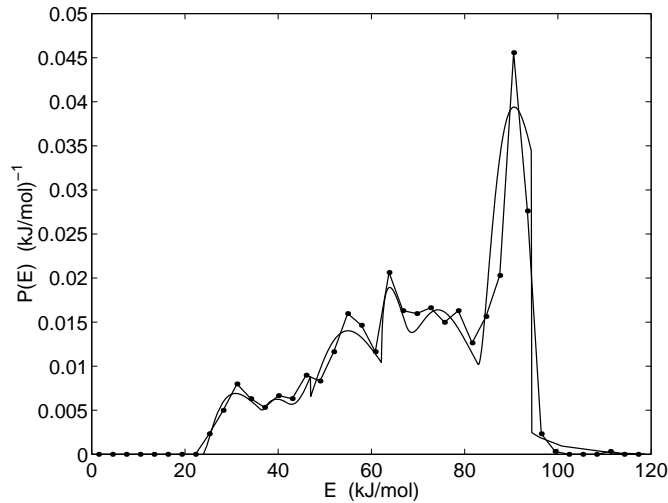}
\caption{Breathers spectra, i.e.,
breather probability densities $P_\br(E)$ obtained numerically
(line with dots) and theoretically (continuous line). The latter
is obtained by considering six different types of breathers. See text.}
 \label{fig:pbe} \end{center}
\end{figure}

Note that the temperature can not be fixed from the outcome. For some energy we obtain
a temperature $T_C=280.25$~C, approximately the experimental one, and the data given
here correspond to that case. We
obtain a mean number of breathers per site $\langle n_\br \rangle =0.92 \times 10^{-3}$,
with a mean energy $\langle E\rangle \sim$ 70 kJ/mol which would correspond to
$z=9.19$ at the experimental temperature. However, the probability density $P_\br (E)$ does
not correspond to a curve as the one given by Eq.~(\ref{eq:pede}) and plotted
in Figure~\ref{fig:pe_brph}, because it is  a curve with several maxima and minima
(see Fig.~\ref{fig:pbe}).
The reason is that with the numerical simulations,
we obtain all types of multibreathers and single breathers with different symmetries
and not only the symmetric, single, exact breathers obtained
numerically in the previous section,  which are continuation
from a single excited oscillator at the
anticontinuous limit.

Therefore, the numerical curve $P_\br (E)$ can be seen as a numerical spectrum
for the different breather energies and forms of vibration. The ideal objective,
as with other types of spectra, would be to know each type of breather, with its
dispersion curve $E(\nu_\br)$, and the relative probability of its appearance and to
be able to reproduce exactly the numerical spectrum.

In principle, all the breather types could be obtained exactly with different
conditions at the anticontinuous limit with different frequencies and by path continuation
by changing the frequency and studying the different branches at the possible
bifurcations. This is a long and difficult task that we do not pursue here. Instead, we
try to fit approximately the numerical $P_\br (E)$, with a small number of breather types,
each one characterized by its minimum energy $\Delta$, maximum energy $E_\mathrm{M}$, which
can be $\infty$, and parameter $z$ (see \ref{subsec:brhard} and \ref{subsec:brsoft}),
each breather type with a different probability
 to occur. In this way, we know that we cannot fit exactly the numerical spectrum
because we are most probably substituting a number of breather types with
an average one. In any case, our numerical $P_\br(E)$ is also an approximation,
as we would need a extremely large number of simulations to obtain the
actual curve.

Note that the breather spectrum will not appear in an experimental
one for three reasons: a)~The number of breathers is about $10^{-3}$ the number of
phonons, and the spectrum is basically dominated by the one-phonon transitions;
b)~Breathers are localized and, therefore, they cannot be excited by infrared, raman
of neutron spectra~\cite{Fillaux04}.

\begin{figure}[ht] \begin{center}
    \includegraphics[clip,width=\singlefig]{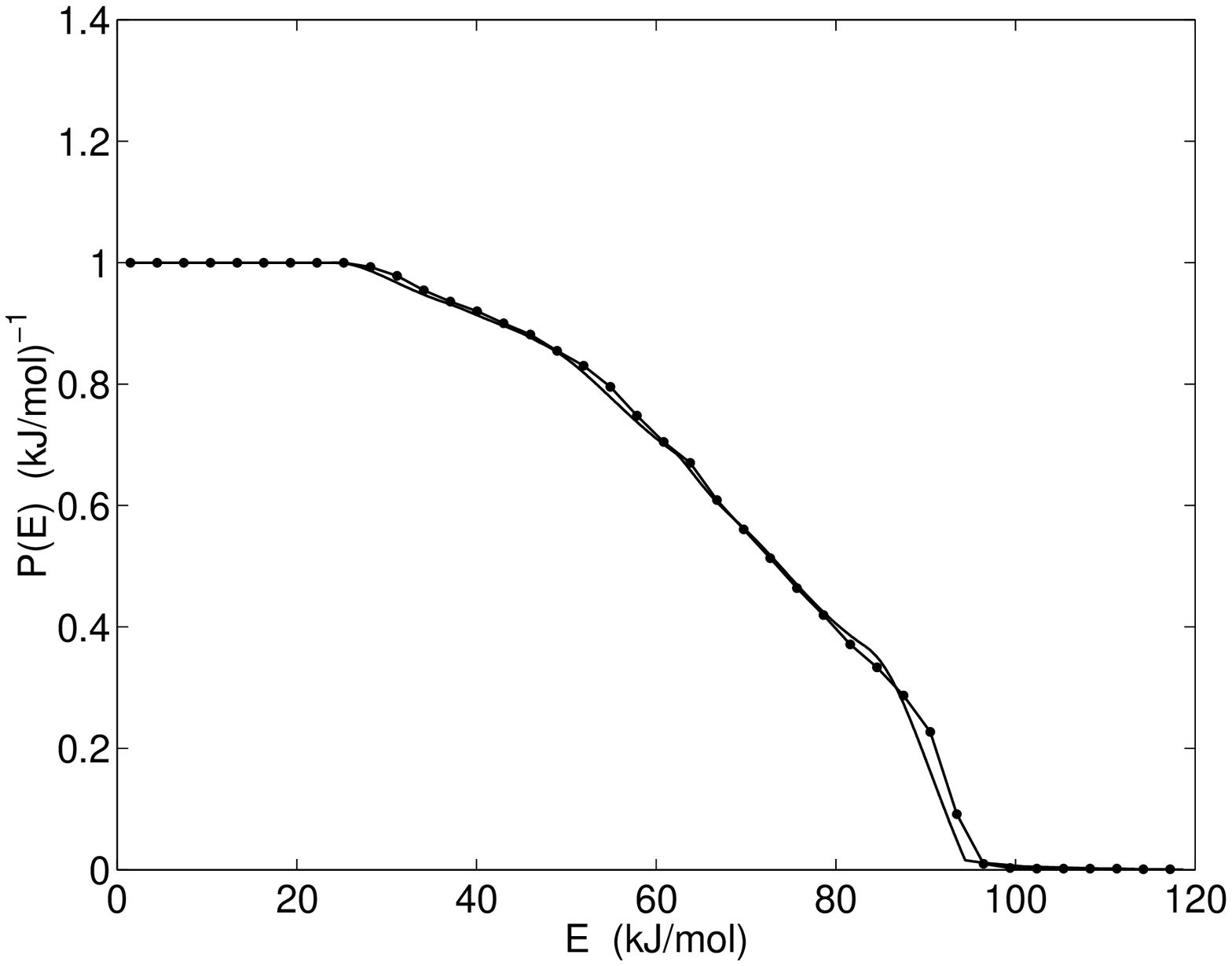}
\caption{Breather cumulative probabilities  $C_\br(E)$ obtained numerically
(line with dots) and theoretically (continuous line). The latter
considering six different types of breathers. See text.}
 \label{fig:cbe} \end{center}
\end{figure}

Figures~\ref{fig:pbe} and \ref{fig:cbe} show the numerical and
theoretical, probability densities
and cumulative probabilities, respectively. The parameters of
the breathers are:

\begin{tabular}{c c c c c c c }
  $\Delta$ (kJ/mol) & 23.9 & 36.6 & 41.4 & 62.2 & 67.3 & 82.9\\
  $z$ & 1.50 & 1.17 & 3.00 & 0.52 & 2.07 & 1.80\\
  $E_\mathrm{M}$ (kJ/mol))& -- & 46.9 & -- & -- & -- & 94.4  \\
  Probability & 0.103 & 0.026 & 0.281 & 0.097 & 0.202 & 0.290 \\
\end{tabular}

\subsection{Effect on the reaction rate}
\label{sub:effect}
According to Ref.~\citenum{Putnis} the lowest estimates of the
activation energy for a reconstructive transformation as the one
described above are $\Ea=100-200\,\mathrm{kJ/mol}$.
Let $n_\mathrm{ph}(\Ea)\simeq \exp(-\beta \Ea)$ and
$n_\br(\Ea)$ be the mean number of phonons and breathers,
respectively, per site, with energies $E\geq \Ea$.
Then, $n_\br(\Ea)=\langle n_\mathrm{b}\rangle C_\br(E_a)$,
with $\langle n_\mathrm{b}\rangle\sim 0.92\cdot 10^{-3}$, the
mean number of breathers per site obtained numerically,
and $C_\br(E_a)$, the cumulative probability described in
the previous subsection, obtained with six different types
of breathers in order to fit the numerical probability density.
The ratio of the number of breathers to phonons
becomes $n_\br (\Ea)/n_\mathrm{ph}(\Ea)\sim 10^4-10^5$.

The reaction rate constant with breathers  would be
$k_\br=A_\br\, n_\br(\Ea)$, with  $A_\mathrm{b}$, the
frequency factor for breathers $A_\mathrm{b}$, which should be different
from $A$, the frequency factor for phonons.
We will assume that $A_\mathrm{b}=A$
for the purpose of comparison.
The ratio of reaction rates for breathers and phonons would be
$k_\br/k=n_\mathrm{b}(\Ea)/n_\mathrm{ph}(\Ea)\sim
10^4-10^5$ for $\Ea=100-200$ kJ/mol. In other words, as
the three days experimental time leads to about 30\% of the
transformation performed, the time without breathers to obtain
the same result, would be $10^4-10^5$ times larger and, thus,
completely unobservable.

Two factors are likely to increase further the reaction rate
with breathers. First, since a discrete breather is strongly localized,
it seems much more capable of delivering the energy for breaking a \mbox{Si-O} bond,
which implies that $A_\mathrm{b}$ should in effect be much
larger than $A$. Second, for larger systems than the one used in our simulations,
the fluctuations may have larger energies and, therefore, excite other
types of breathers with higher energies, as, for example, breathers with
frequencies above the phonon band, which have energies between 240
and 500~kJ/mol. These breathers
would increase the fraction of breathers above the
activation energy and therefore the reaction rate.

\section{Summary and conclusions} Low temperature reconstructive
transformations (LTRT) have been achieved in layered silicated by
some of the authors at temperatures about 600$^\circ$C lower than
previously reported. This is a phenomenon for which there is
presently no plausible  explanation since the bonds involved are the same
as in other transformations. New experiments performed by some of
the authors on mica muscovite, a non-expandable silicate have
discarded their previous hypothesis of LTRT been caused by the
expansion of the intersheet layer.

We have constructed a model for breathers in the cation later, for
which we have obtained reasonable parameters, and with a mixture of
numerics and theory we have estimated their effect in the reaction
rate. The results are that they would increase enormously the
reaction rate and, thus, explain the observed LTRT. This can be
easily explained by the fact that, although there are much less
breathers than phonons, there are many more with energies above
the expected activation
energy. Certainly, the statistical theory of breathers is only an
approximation, and the numerics cannot be precise at larger energies
for which there are so few breathers, except if an enormous number
of simulations could be performed. However, an established fact in
breather theory is that large breathers have longer life time than
small ones and thus tend to overpopulate the regions of high
energies if compared to Maxwell-Boltzmann statistics for phonons.
Moreover, as they are localized, it seems that they can deliver more
easily the required energy to break bonds. The sum of this facts,
i.e., localization, much higher number of breathers above a given
activation energy and apparent diminution of the activation energy
suggests that DBs are good candidates to explain LTRTs.

\section*{Acknowledgments}
JFRA and JC acknowledge sponsorship by the Ministerio de
Educaci\'on y Ciencia, Spain, project FIS2004-01183. MDA, MN and
JMT acknowledge sponsorship from the same body, projects
MAT2002-03504 and CTQ2004-05113. JFRA acknowledges the
hospitality and the spectra performed at CNRS-LADIR. All the
authors acknowledge Prof. R. Livi, from Florence University for
useful discussions.

\section*{Appendices}
\appendix

\renewcommand{\theequation}{A-\arabic{equation}}
  \setcounter{equation}{0}  
\section{Phonon statistics}
\label{ap:phonon}
\subsection{Quantum statistics of one oscillator}
\label{subsec:1oscillator}
A quantum harmonic oscillator with frequency $\w$, equal to its
classical one, has energies $E_n=(n+\frac{1}{2})\,\hbar\,\w$.
If in contact with a thermal bath at temperature $T$, the probability that it has
energy $E_n$ is given by $P(E_n)=A\,\exp(-\beta\,E_n)=A\exp(-\beta\,(n+\frac{1}{2})\,\hbar\,\w)$,
with $\beta=1/\mathrm{k}_\mathrm{B} T$, $\mathrm{k}_\mathrm{B}$ being the Boltzmann constant.

A can be obtained by the normalization condition $\sum_{n=0}^\infty P_n=A\,Z=1$, where
$Z=\sum_{n=0}^\infty \exp[-\beta\,(n+\frac{1}{2})\,\hw]$ is known as the partition function for the
oscillator. $Z$ is a geometric series which can be easily summed leading to:
\begin{equation}
             Z=\frac{\exp(-\beta\fracc{\hw}{2})}{1-\exp(-\beta\hbar\w)}.
             \label{eq:Z}
\end{equation}
Therefore $A=1/Z$ and
\begin{equation}
             P_n\equiv P(E_n)=\frac{\exp[-\beta (n+\frac{1}{2})\hbar\w]}{Z}=
             \exp(-\beta n \hw)[1-\exp(-\bhw)]
             \label{eq:PE}
\end{equation}

Note that $n$ is the excitation number of the oscillator, however, in solid state physics,
where $\w$ is the frequency of a normal mode of the solid, it is customary to speak of $n$
as the number of phonons with frequency $\w$.
Once known $P_n$ a number of quantities can be readily calculated.
The mean energy $\langle E \rangle $=$Z^{-1} \sum_{n=0}^{\infty}(n+\frac{1}{2})\hw
\exp[-\beta (n+\frac{1}{2})\hbar\w]$ =
$-Z ^{-1}\partial Z/\partial \beta$ = $-\partial \log(Z)/\partial \beta$, which leads to:
\begin{equation}
             \langle E \rangle =\big(\frac{1}{2}+\frac{1}{\exp(\beta\hbar\w)-1}\big)\hw\,.
             \label{eq:meanE}
\end{equation}

Consequently, the mean excitation number or mean number of phonons is:
\begin{equation}
             \langle n \rangle =\frac{1}{\exp(\beta\hbar\w)-1}.
             \label{eq:meann}
\end{equation}

For high temperatures $\hbar\w/\kb T<<1$, the mean energy
becomes $\langle E \rangle \simeq \kb T$ which is the classical mean energy of a
harmonic oscillator and $\langle n \rangle \simeq \kb T/\hbar \w$.

Of particular importance for the present problem is the cumulative probability $C(\Ea)$,
i.e., the probability
that the oscillator
has energy $\Ea$ or higher above the ground state $\hbar \w/2$, i.e., that it can deliver
the energy $\Ea$. Let $m$ be the minimum
integer so as $m\,\hbar \w\geq \Ea$, i.e., $m=\lceil \Ea/\hbar\w\rceil$, with
 $\lceil x \rceil$, the ceiling function that rounds $x$ towards
plus infinity. Then
$C(\Ea)$ = $Z^{-1}\sum_{n=m}^\infty \exp[-\beta (\frac{1}{2}+n)\hbar \w]$ =
$Z^{-1}\sum_{n=0}^\infty \exp[-\beta (m+\frac{1}{2}+n) \hbar \w]$ =
$Z^{-1}\exp(-\beta m \hbar\w)$ $\times$ $\sum_{n=0}^\infty \exp[-\beta (\frac{1}{2}+n)\hbar \w]$,
therefore:
\begin{equation}
         C(\Ea)=
         \exp(-\beta m \hbar\w)=
          \exp(-\beta \lceil \frac{ \Ea}{\hw}\rceil \hbar\w)\,.
             \label{eq:Pa}
\end{equation}
If $m$ is of the order of a few tens, the expression above approaches to
the classical expression $C_{\mathrm{class}}(\Ea)=\exp(-\beta \Ea)$.
 Note, however, that the quantum probability is
  somewhat smaller. The ratio between the quantum probability and
  the classical one is between $\exp(-\beta \hbar \w)$ and 1.
  For a frequency as the one given here for the phonons in the K$^{+}$
  plane $\omega_0=3.16 \cdot 10^{13}\mathrm{s}^{-1}$ and $T=573 \mathrm{K}$ this ratio
is $\exp(-\beta \hbar \,\omega_0)\approx 0.65$ and $0.81$ for $T=1173 \mathrm{K}$.

\subsection{Normal modes and phonons}

Let us consider a solid in the linear approximation, with $N_f$ degrees of freedom, with
$N_f=3\times N_a$ for a three-dimensional solid with $N_a$ atoms.
There are $N_f$ normal modes with frequencies
$\w_i$ and wave numbers $\bf{k}_i$, each one equivalent to a independent harmonic oscillator,
and therefore, the previous subsection can be applied to it.
The properties of the solid are simply the sum of the properties of the isolated
oscillators, just adding the subindex $i$ to the formulae in the preceding section and summing up,
 i.e.,
the excitation number $\langle n_i \rangle $
of the mode $i$ (or the number of phonons)  and the energy of the solid are given by:
\begin{equation}
\langle n_i \rangle=\frac{1}{\exp (\beta \hbar \w_i)+1}\quad ;\quad
E=\sum_{i=1}^{N_f} (\langle n_i \rangle+\frac{1}{2})\hbar \w_i
\end{equation}

The number of modes  $N_{ph}(\Ea)$ with energy larger or equal to $\Ea$ above
their ground state is
\begin{equation}
 N_{ph}(E\geq \Ea)=\sum_{i=1}^{N_f} \exp(-\beta \lceil
 \frac{\Ea}{\hbar \w_i} \rceil \hbar \w_i) \, .
\end{equation}
 If $\Ea$ is
about a few tens larger than any $\hbar \w_i$, then we simply have:
\begin{equation}
 N_{ph}(E\geq \Ea)\simeq N_f \exp(-\beta \Ea) \, ,
\end{equation}
which is the classical expression.
Again, the ratio between the quantum and classical expressions of $N_{ph}(E\geq \Ea)$
is somewhat smaller than the unity.

Therefore, the cumulative probability $C(\Ea)$ i.e., the fraction of
modes with energies greater of equal to $\Ea$, becomes:
\begin{equation}
 C(\Ea)\simeq \exp(-\beta \Ea) \, .
 \label{eq:cumprobphon}
\end{equation}

\section{Breather statistics}
\label{ap:breather}
\renewcommand{\theequation}{B-\arabic{equation}}
  \setcounter{equation}{0}  
\subsection{Breathers with hard on-site potential}
\label{subsec:brhard}
The breather statistics theory developed in Ref.~\citenum{Piazza03} for 2D breather
in a system with hard on--site potential is based
 is some simple hypotheses, which, in principle, can be fairly general:

\begin{enumerate}
\item An established fact is that
breathers in two and three dimensions have a minimum energy $\Delta$~\cite{FKM97}.

\item The rate of creation of breathers with energy $E$, $B(E)$,
is proportional to $\exp(-\beta E)$,
since breathers form from fluctuations through an activation process.

\item The probability per unit time that a breather with energy $E$ is destroyed,
 $D(E)$, is postulated to be inversely proportional to
 $(E-\Delta)^{z}$, with $z$ a constant (which means
as other constants hereafter that it does not change with the energy $E$) that
 depends on the system.
 This  law is the simplest mathematical expression that
  takes into account that large breathers have longer lives
 than smaller ones, with  the lifetime
of breathers with minimum energy $\Delta$ equal to zero. It has to
be considered as an approximation as it is not derived from first
principles.
\end{enumerate}

Let $P_\br (E)\,\d E$ be the probability of existence (or the mean fraction)
of breathers with energy between $E$ and $E +\d E$.
The rate of destruction of breathers with
energy $E$ is proportional to $D(E)$ and $P_\br(E)$, therefore,
$\exp(-\beta E)$ =
$A\,P_\br(E)(E-\Delta)^{-z}$, $A$ being a constant, or,
$A\,P_\br (E)$=$(E-\Delta)^z \exp(-\beta E)$.
Since $\int_\Delta^\infty P_\br (E)\d E=1$, $A$ can be obtained
 using the change of variable $y$ =
$\beta (E-\Delta)$:
\begin{eqnarray}
\lefteqn{ A=\int_\Delta^\infty (E-\Delta)^z\exp(-\beta E)=}
 \nonumber \\
& &\frac{\exp(-\beta \Delta)}{\beta^{z+1}}\int_0^\infty y^z\exp(-y)\d y =
\exp(-\beta \Delta)\beta^{-(z+1)}\Gamma(z+1)\,,\label{eq:A}
\end{eqnarray}
with $\Gamma(z+1)=\int_0^\infty y^z\exp(-y)\d y$, the Gamma function. Thus, the
probability of breathers with energy between $E$ and $E+\d E$ is given by:
\begin{eqnarray}
P_\br (E)=\frac{1}{A}(E-\Delta)^z \exp(-\beta E)=
\frac{(E-\Delta)^z \exp(-\beta E)}{\exp(-\beta \Delta)\beta^{-(z+1)}\Gamma(z+1)}=
\nonumber\\
\frac{\beta^{z+1}}{\Gamma(z+1)}(E-\Delta)^z\exp[-\beta (E-\Delta) ]\,.
\label{eq:P_bE}
\end{eqnarray}

The mean energy is given by:
\begin{eqnarray}
\langle E \rangle =\int_\Delta^\infty E\, P_\br(E)\d E=\Delta +\int_\Delta^\infty (E-\Delta)
P_\br(E)\d E=
\nonumber \\ \Delta+\int_\Delta^\infty
\frac{\beta^{z+1}}{\Gamma(z+1)}(E-\Delta)^{z+1}\exp[-\beta (E-\Delta) ]\d E=
      \nonumber \\
      \Delta+  \frac{1}{\beta \,\Gamma(z+1)}\int_0^\infty y^{z+1}\exp(-y) \d E =
\Delta+\frac{\Gamma(z+2)}{\beta \,\Gamma(z+1)}=\nonumber \\
\Delta+(z+1)\,\kb T\,.
\label{eq:Emb}
       \end{eqnarray}

The cumulative probability $C_\br(E)$, i.e., the probability that a breather has
energy higher than $E$, is given by:
\begin{eqnarray}
C_\br(E)&=&\int_E^\infty P_\br(E)\d E=\int_E^\infty
\frac{\beta^{z+1}}{\Gamma(z+1)}(E-\Delta)^z\exp[-\beta (E-\Delta) ]\,\d E=\nonumber \\
    & &   \frac{1}{\Gamma(z+1)}\int_{\beta (E-\Delta)}^\infty y^z \exp(-y) \d y =
\frac{\Gamma(z+1,\beta (E-\Delta))}{\Gamma(z+1)}\,,
\label{eq:CbE}
\end{eqnarray}
where $\Gamma (z+1,x)$ = $\int_x^\infty y^z \exp (-y) \d y$ is
the first incomplete Gamma function~\cite{Abram}.

\begin{figure}[h] \begin{center}
   \includegraphics[clip,width=\doublefig]{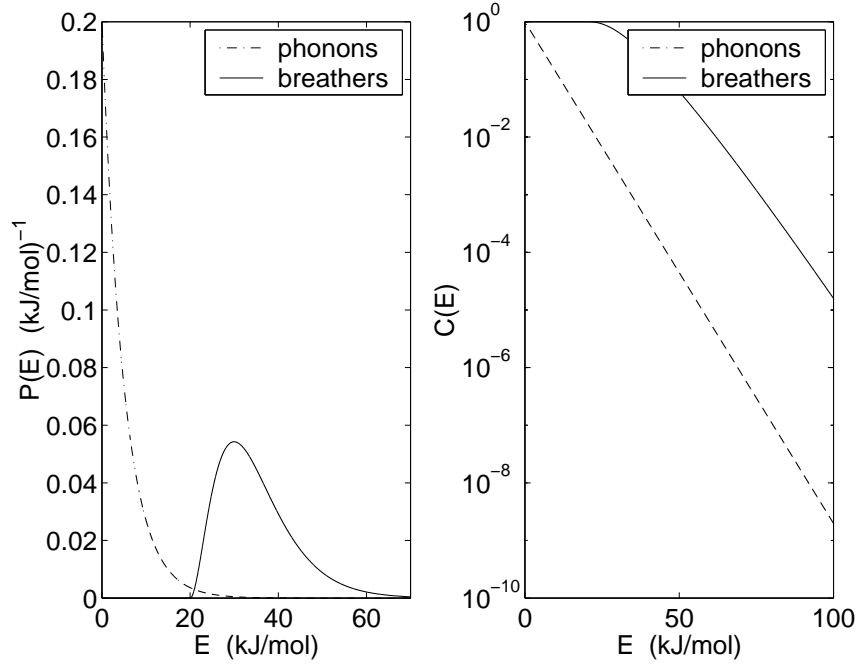}
   \caption{Comparison of the phonon and breather probabilities densities (left)
   and cumulative probabilities (right).
   The breather values have been obtained for $z=2$ and $\Delta=20$~kJ/mol.
The temperature is $T=600~$K and $\kb T\approx 5$~kJ/mol.}
    \label{fig:pe_brph}
    \end{center}
\end{figure}

The energy for which the probability is maximum is given by $E(P_{\rm{max}})=\Delta +z\,\kb\,T$,
which shows that breathers tend to populate higher energies than phonons. 
As an example, Fig.~\ref{fig:pe_brph} shows
$P_\br (E)$ and $C_\br(E)$ for breathers with $\Delta=20$~kJ/mol and $z=2$
and the equivalent magnitudes for phonons.
The large energies the larger values of $C_\br(E)$ soon compensate for
the much smaller number of breathers than phonons (around $10^{-3}$).
Although the extrapolation of
$C_\br(E)$ to large energies has to be done with caution and
a more elaborate theory has yet to be developed,
the basic fact that breathers tend to populate higher energies than phonons
can be accepted.
Piazza et al~\cite{Piazza03} have succeeded in fitting the cumulative probability
in Eq.~\ref{eq:CbE}
with the observed one in numerical experiments, which proves that the hypotheses
above are reasonable.

\subsection{Breathers with maximum energy}
\label{subsec:brsoft}
Hereafter we modify slightly their theory developed above. The on-site potential
for the 2D system in Ref.~\citenum{Piazza03}
 is hard, i.e., the frequency of the isolated oscillators
increases with the frequency, with the consequences that the breather frequency lies
above the phonon band, its energy increases with its frequency and can be
considered as unbounded.

 For soft on-site potentials or, as in the present paper,
 potentials with both soft and hard parts, breather energies may have an upper limit
 because the breather frequency enters the phonon band or because a bifurcation
 where the breather disappears or transforms into a different one, as a multibreather or
 a breather with different symmetries. The changes are obtained by introducing an
 upper limit $E_\mathrm{M}$ in the integrals with respect to the energy. Then, Eq.~(\ref{eq:A})
 becomes (with $y=\beta\,(E-\Delta)$):
\begin{eqnarray}
A=\int_\Delta^{E_\mathrm{M}} (E-\Delta)^z\exp(-\beta E)=
\frac{\exp(\beta \Delta)}{\beta^{z+1}}\int_0^{\beta\,(E_\mathrm{M}-\Delta)} y^z\exp(-y)\d y=\nonumber \\
 \exp(\beta \Delta)\beta^{-(z+1)}\gamma(z+1,\beta\,(E_\mathrm{M}-\Delta))\,,\label{eq:Am}
\end{eqnarray}
where $\gamma(z+1,x)=\int_0^x y^z\exp(-y)\,\d y$ is the second incomplete gamma
 function~\cite{Abram}.

 Therefore, the probability density becomes:
\begin{eqnarray}
P_\br (E)=\frac{1}{A}(E-\Delta)^z \exp(-\beta E)=
\frac{\beta^{z+1}(E-\Delta)^z\exp[-\beta (E-\Delta)]}{\gamma(z+1,\beta\,(E_\mathrm{M}-\Delta))} \,.
\label{eq:P_bEm}
\end{eqnarray}
The cumulative probability becomes:
\begin{eqnarray}
C_\br (E)=\int_E^{E_\mathrm{M}} P_\br(E)\,\d E=
\int_E^{E_\mathrm{M}} \frac{\beta^{z+1}(E-\Delta)^z\exp[-\beta (E-\Delta)]}
{\gamma(z+1,\beta\,(E_\mathrm{M}-\Delta))}\,\d E=\nonumber \\
      \frac{1}{\gamma(z+1,\beta(E_\mathrm{M}-\Delta))}
       \int_{\beta (E-\Delta)}^{\beta (E_\mathrm{M}-\Delta)} y^z \exp(-y) \d y =
1-\frac{\gamma(z+1,\beta (E-\Delta))}{\gamma(z+1,\beta (E_\mathrm{M}-\Delta))}\,.
\label{eq:CbEm}
\end{eqnarray}
For $E_\mathrm{M}>>\Delta$, the expressions above for $P_\br(E)$ and  $C_\br(E)$
transform into
the expressions in Eqs.~(\ref{eq:Emb},\ref{eq:CbE})

In a system like ours there are different types of breathers and
the probabilities or cumulative probabilities calculated
above correspond
to each type  with different minimum and maximum energies
$\Delta$ and $E_\mathrm{M}$ (or without maximum energy), and parameter $z$.
The total number of breathers and
the relative probability of each type are unsolved questions. The
latter probably depends on the temperature, the breather energies, the phase space
occupied by each breather and its equivalent ones through symmetries, and
the breather profile, which might be excited more or less easily by the
phonons.


\newcommand{\noopsort}[1]{} \newcommand{\printfirst}[2]{#1}
  \newcommand{\singleletter}[1]{#1} \newcommand{\switchargs}[2]{#2#1}
\providecommand{\refin}[1]{\\ \textbf{Referenced in:} #1}

\end{document}